\documentclass{article}
\usepackage{spconf,amsmath,epsfig}
\usepackage{algorithm}
\usepackage{subcaption}
\usepackage{bm}
\usepackage{footnote}
\usepackage{upgreek}
\usepackage{epsf,exscale,times}
\usepackage{graphicx}
\usepackage{amssymb}
\usepackage{amsmath}
\usepackage{amsthm}
\usepackage{amsfonts}
\usepackage{fancyhdr}
\usepackage{tikz}
\usepackage{multirow}
\usepackage{multicol}
\usepackage{glossaries}
\usepackage{cite}
\usepackage{nccmath}
\usepackage{commath}
\usepackage{comment}
\usepackage{color}
\usepackage{mathtools}
\usepackage{booktabs}
\usepackage{siunitx}
\usepackage{hyperref}
\usepackage{cuted}
\usepackage{xurl}
\usepackage{subcaption}

\newacronym{APAS}{APAS}{Almost Perfect auto-correlation Sequences}
\newacronym{MPS}{MPS}{Minimum Peak Sidelobe} 
\newacronym{PMCW}{PMCW}{Phase Modulated Continuous Wave}
\newacronym{MF}{MF}{Matched Filter}
\newacronym{SNR}{SNR}{Signal to Noise Ratio}
\newacronym{INR}{INR}{Interference to Noise Ratio}
\newacronym{SINR}{SINR}{Signal to Interference plus Noise Ratio}
\newacronym{AF}{AF}{Ambiguity Function}
\newacronym{MIMO}{MIMO}{Multiple Input Multiple Output}
\newacronym{SISO}{SISO}{Single Input Single Output}
\newacronym{CD}{CD}{Coordinate Descent}
\newacronym{BCD}{BCD}{Block Coordinate Descent}
\newacronym{GD}{GD}{Gradient Descent}
\newacronym{MM}{MM}{Majorization-Minimization}
\newacronym{FMCW}{FMCW}{Frequency Modulated Continuous Wave}
\newacronym{CDM}{CDM}{Code Division Multiplexing}
\newacronym{DFT}{DFT}{Discrete Fourier Transform}
\newacronym{FFT}{FFT}{Fast Fourier Transform}
\newacronym{MVDR}{MVDR}{Minimum Variance Distortionless Response}
\newacronym{MBI}{MBI}{Maximum Block Improvement}
\newacronym{RFPA}{RFPA}{Radio Frequency Power Amplifier}
\newacronym{BPSK}{BPSK}{Binary Phase Shift Keying}
\newacronym{QPSK}{QPSK}{Quadrature Phase Shift Keying}
\newacronym{ULA}{ULA}{Uniform Linear Array}
\newacronym{DOF}{DOF}{Degrees of Freedom}
\newacronym{PSK}{PSK}{Phase Shift Keying}
\newacronym{PSL}{PSL}{Peak Sidelobe Level}
\newacronym{ISL}{ISL}{Integrated Sidelobe Level}
\newacronym{ISLR}{ISLR}{Integrated Sidelobe Level Ratio}
\newacronym{SILR}{SILR}{Spectral Integrated Level Ratio}
\newacronym{LFM}{LFM}{Linear Frequency Modulation}
\newacronym{HPM}{HPM}{Hybrid Phased MIMO}
\newacronym{MPSK}{MPSK}{$M$-ary Phase Shift Keying}
\newacronym{LPI}{LPI}{Low Probability of Intercept}
\newacronym{RoC}{RoC}{Radar-on-Chip}
\newacronym{RF}{RF}{Radio-Frequency}
\newacronym{PAR}{PAR}{Peak-to-Average Power Ratio}
\newacronym{LTE}{LTE}{Long Term Evolution}
\newacronym{DL}{DL}{Down Link}
\newacronym{UL}{UL}{Up Link}
\newacronym{iid}{i.i.d.}{independent and identically distributed}
\newacronym{BS}{BS}{Base Station}
\newacronym{BSUM}{BSUM}{Block Successive Upper-bound Minimization}
\newacronym{SDR}{SDR}{Software-Defined Radio}
\newacronym{OTA}{OTA}{Over-The-Air}
\newacronym{USRP}{USRP}{Universal Software Radio Peripheral}
\newacronym{ICCL}{ICCL}{Integrated Cross Correlation Level}
\newacronym{ADMM}{ADMM}{Alternating Direction Method of Multipliers}
\newacronym{SDPM}{SDPM}{Spectrum Discrete Phase Modulation}
\newacronym{CW}{CW}{Continuous Wave}
\newacronym{DoA}{DoA}{Direction of Arrival}
\newacronym{MUSIC}{MUSIC}{Multiple Signal Classification}
\newacronym{BiST}{BiST}{Binary Sequences seTs}
\newacronym{PDSCH}{PDSCH}{Physical Downlink Shared Channel}
\newacronym{PDCCH}{PDCCH}{Physical Downlink Control Channel}
\newacronym{MCS}{MCS}{Modulation and Coding Schemes}
\newacronym{GUI}{GUI}{Graphical User Interface}
\newacronym{MI}{MI}{mutual information}
\newacronym{NI}{NI}{National Instruments}
\newacronym{HW}{HW}{hardware}
\newacronym{ADC}{ADC}{Analog-to-digital converter}
\newacronym{LS}{LS}{Least Squares}
\newacronym{MSE}{MSE}{mean-squared error}
\newacronym{ISNR}{ISNR}{Inverse Signal to Noise Ratio}
\newacronym{SOCP}{SOCP}{Second-Order
Cone Programming}
\newacronym{QCQP}{QCQP}{Quadratically constrained
quadratic program}
\title{Improving Pulse-Compression Weather Radar via the Joint Design of Subpulses and Extended Mismatch Filter}
\name{Linlong~Wu, Mohammad~Alaee-Kerahroodi, and Bhavani Shankar~M. R.}
\address{Interdisciplinary Center of Security, Reliability and Trust (SnT), University of Luxembourg\\
Email:\{linlong.wu, mohammad.alaee,bhavani.shankar\}@uni.lu}
\begin{document}
%
\maketitle
\begin{abstract}
Pulse compression can enhance both the performance in range resolution and sensitivity for weather radar. However, it will introduce the issue of high sidelobes if not delicately implemented. Motivated by this fact, we focus on the pulse compression design for weather radar in this paper. Specifically, we jointly design both the subpulse codes and extended mismatch filter based on the alternating direction method of multipliers (ADMM). This joint design will yield a pulse compression with low sidelobes, which equivalently implies a high signal-to-interference-plus-noise ratio (SINR) and a low estimation error on meteorological reflectivity.
The experiment results demonstrate the efficacy of the proposed pulse compression strategy since its achieved meteorological reflectivity estimations are highly similar to the ground truth.
\end{abstract}
\section{Introduction}
\label{sec:intro}

In a weather radar system, sensitivity and resolution are the two important performance metrics. The sensitivity, referring to the minimum detectable reflectivity, is inversely proportional to the product of the pulse width and peak power \cite{bc}. Solid-state transmitters with high peak power are usually difficulty or expensive to obtain for most meteorological applications. Consequently, longer pulses must be transmitted to attain adequate sensitivity. 
However, the range resolution is proportional to the pulse width, and thereby longer pulses will results in the resolution degeneration. 
To tackle this dilemma, pulse compression can be the remedy for a low peak-power, long-duration coded pulse system to attain both the fine range resolution and improved sensitivity. 

As a signal processing technique, pulse compression widens the bandwidth of the transmitted pulse by modulating it in either phase or frequency, and then the received echo is usually processed by a matched filter which compresses the long pulse to a short duration. 
The first attempt of implementing pulse compression on weather radar systems is back to 1970s \cite{fetter1970radar}, and some works investigated several technical and engineering aspects, see \cite{mudukutore1998pulse,george2008considerations,beauchamp2016pulse} and references therein. However, comparing to the wide applications in non-weather radars, its application on weather radar is relatively rare. The main obstacle lies in the range sidelobes introduced by the pulse compression, which is more notorious for weather radar since the targets of interest are usually extended volume scatters \cite{kumar2020intrapulse}. A large sidelobe will eventually cause inaccurate estimation of the reflectivity amplitude, which is a key index in meteorological applications. To tackle the issue of sidelobe, many works have been conducted recently \cite{song2015optimization,kerahroodi2017coordinate} although they were not primarily for the weather radar. In those works, the transmitted waveform is delicately designed to achieve a low peak or integrated sidelobe level. 

For weather radar, benefiting from these developed techniques, the outlook towards using pulse compression could become promising after some appropriate adaptions. Motivated by this insight, in this paper, we investigate the joint design of transmit subpulses and mismatched filter for pulse compression weather radar. In general, we adopt the extended mismatched filter strategy with zero padding to improve sidelobes. Then we formulate the optimization problem and show the close relation among SINR, sidelobe and estimation error. To solve this nonconvex problem, we propose an efficient method based on the alternating direction method of multipliers (ADMM) \cite{boyd2011distributed}. The experiment results shows that the efficacy of the proposed pulse compression strategy and the enhanced performance. 

\section{Problem Formulation}
\label{sec:format}

Let $\boldsymbol{J}_k$ be a $N \times N$ shift matrix with the $(m,n)$-th entry being
\begin{equation}\label{eq:Jk}
\boldsymbol{J}_{k}\left(m,\text{ }n\right)=\begin{cases}
1,\text{ } & m-n= k \\
0,\text{ } & m-n\ne k.
\end{cases}
\end{equation}
Also, let $\boldsymbol{x} = [x_1, x_2, \ldots, x_N]^T \in {\mathbb{C}}^N$ be the transmitted fast-time radar code vector. 
Then, the received signal 
$\boldsymbol{y} \in \mathbb{C}^{N}$
after sampling from the range bin of interest is \cite{4567600},
\begin{equation}\label{eq:rcv}
    \boldsymbol{y} =  \alpha_0 \boldsymbol{x} +
    {\sum_{\substack{k = -N+1, k\neq 0}}^{N-1} \alpha_k \boldsymbol{J}_k \boldsymbol{x}} + \boldsymbol{\nu},
\end{equation}
where $\alpha_k,\forall k=-N+1,\ldots,N+1$ is a complex-valued scalar of atmospheric reflectivity corresponding to the $k$-th range bin illuminated by the radar pulse compression code, which is assumed to be independent and
\begin{equation}
    \mathbb{E}\{|\alpha_k|^2\} = \zeta_k, \forall k=-N+1,\ldots,N+1,
\end{equation}
and $\boldsymbol{\nu}\sim \mathcal{N}\left(\boldsymbol{0},\sigma_{\nu}^{2}\boldsymbol{I}\right)$ is the noise vector which is uncorrelated with the other signal-dependent terms. 

One important step in weather radar is to estimate $\alpha_0$, which implies the hydrometer phenomena and is proportional to the reflectivity. In the conventional weather radars without pulse compression, the received signal in \eqref{eq:rcv} does not contain the second term. Therefore, when using the pulse compression for a refined range resolution and lower peak power level, minimizing the introduced interference becomes critical.

By deploying mismatched-filter-like techniques, we can enlarge the filter length compared to the length of the subpulses after zero padding, which will spread the energy of the auto-correlation sidelobes into much larger coefficient space \cite{kumar2020intrapulse}.
The subpulses after zero padding is defined by
\begin{equation}
    \widetilde{\boldsymbol{x}} = [\boldsymbol{0}^T_M, \boldsymbol{x}^T, \boldsymbol{0}^T_M]^T,
\end{equation}
and let
$\widetilde{\boldsymbol{w}} \in \mathbb{C}^{\widetilde{N}}$ be the receive filter with $\widetilde{N} = 2 M + N$. 
The received signal after filtering can be obtained by
\begin{equation}\label{eq:rcvw}
    \widetilde{\boldsymbol{z}} = \alpha_0 \widetilde{\boldsymbol{w}}^H\widetilde{\boldsymbol{x}} +
    \sum_{\substack{k = -\widetilde{N}+1, k\neq 0}}^{\widetilde{N}-1} \alpha_k \widetilde{\boldsymbol{w}}^H\widetilde{\boldsymbol{J}}_k \widetilde{\boldsymbol{x}} + \widetilde{\boldsymbol{w}}^H\widetilde{\boldsymbol{\nu}},
\end{equation}
where 
$\widetilde{\boldsymbol{J}}_k$ is $\widetilde{N}  \times \widetilde{N}$ shift matrix with the $(m,n)$-th entry being $1$ for $m-n = k$ and $0$ otherwise, and $\widetilde{\boldsymbol{\nu}}\sim \mathcal{N}\left(\boldsymbol{0},\sigma_{\widetilde{\nu}}^{2}\boldsymbol{I}\right)$.

The  instrumental variable  estimate of $\alpha_0$ is given by \cite{4567600},
\begin{equation} \label{eq:est_alphaMMF}
    \hat{\alpha}_0 = \frac{\widetilde{\boldsymbol{z}}}{\widetilde{\boldsymbol{w}}^H\widetilde{\boldsymbol{x}}},
\end{equation}
and its mean square error (MSE) is gievn by
\begin{equation}\label{eq:MSEMMF}
    \text{MSE}(\hat{\alpha}_0) = \mathbb{E}\left\{ \left| \frac{\widetilde{\boldsymbol{z}}}{\widetilde{\boldsymbol{w}}^H\widetilde{\boldsymbol{x}}} - \alpha_0 \right|^2\right\} = \frac{\widetilde{\boldsymbol{w}}^H \boldsymbol{R} \widetilde{\boldsymbol{w}} }{ |\widetilde{\boldsymbol{w}}^H\widetilde{\boldsymbol{x}}|^2},
\end{equation}
where $\boldsymbol{R} = \sum_{\substack{k = -\widetilde{N}+1, k\neq 0}}^{\widetilde{N}-1} \zeta_k  \widetilde{\boldsymbol{J}}_k \widetilde{\boldsymbol{x}} \widetilde{\boldsymbol{x}}^H \widetilde{\boldsymbol{J}}_k^H + \sigma_{\widetilde{\nu}}^2 \boldsymbol{I}$.

In addition, the received SINR for \eqref{eq:rcvw} can be expressed by
\begin{equation}\label{eq:SINRw}
    \text{SINR} = \frac{\zeta_0 |\widetilde{\boldsymbol{w}}^H\widetilde{\boldsymbol{x}}|^2}{\widetilde{\boldsymbol{w}}^H \boldsymbol{R} \widetilde{\boldsymbol{w}} }.
\end{equation}
We can see that maximizing the SINR is equivalent to minimizing the above MSE. Further, $\widetilde{\boldsymbol{w}}^H \boldsymbol{R} \widetilde{\boldsymbol{w}}$ represents the sidelobe interference (ignoring the constant $\widetilde{\boldsymbol{w}}^H\widetilde{\boldsymbol{w}}$). Thus, maxmizing the SINR is essentially maximizing the signal-to-sidelobe ratio, which is desired in pulse compression design.

Based on the illustration, the problem is formulated as
\begin{equation}
\label{eq:10}
\begin{aligned} & \underset{\boldsymbol{x},\widetilde{\boldsymbol{w}}}{\text{minimize}} &  & \frac{\widetilde{\boldsymbol{w}}^H \boldsymbol{R} \widetilde{\boldsymbol{w}} }{ |\widetilde{\boldsymbol{w}}^H\widetilde{\boldsymbol{x}}|^2}\\
 & \text{subject to} &  & \left|x_{i}\right|=1.
\end{aligned}
\end{equation}
Once the optimal $\widetilde{\boldsymbol{w}}$ and $\boldsymbol{x}$ are obtained, it is expected that the estimation of $\alpha_0$ will be enhanced by deploying the designed interpulses $\boldsymbol{x}$ and mismatched filter $\widetilde{\boldsymbol{w}}$.

\section{Proposed Optimization Based Approach}
\label{sec:pagestyle}
In this section, we will focus on solving problem \eqref{eq:10}. 
The alternating optimization (AO) \cite{4787625} will be deployed. 
For a fixed $\boldsymbol{x}_t$ at the $t$-th iteration of AO, the problem w.r.t $\widetilde{\boldsymbol{w}}$
is 
\begin{equation}
\begin{aligned} & \underset{\widetilde{\boldsymbol{w}}}{\text{minimize}} &  & \frac{\widetilde{\boldsymbol{w}}^{H}\boldsymbol{R}_t\widetilde{\boldsymbol{w}}}{\widetilde{\boldsymbol{w}}^{H}\widetilde{\boldsymbol{x}}_{t}\widetilde{\boldsymbol{x}}^{H}_{t}\widetilde{\boldsymbol{w}}}\end{aligned}
\end{equation}
with $\boldsymbol{R}_t = \sum_{\substack{k = -\widetilde{N}+1, k\neq 0}}^{\widetilde{N}-1} \zeta_k  \widetilde{\boldsymbol{J}}_k \widetilde{\boldsymbol{x}}_t \widetilde{\boldsymbol{x}}^H_t \widetilde{\boldsymbol{J}}_k^H + \sigma_{\widetilde{\nu}}^2 \boldsymbol{I}$. It has a closed form solution $\widetilde{\boldsymbol{w}}=\boldsymbol{R}^{-1}_t\widetilde{\boldsymbol{x}}_t$.

For a fixed $\widetilde{\boldsymbol{w}}_{t+1}$, the problem w.r.t $\boldsymbol{x}$ can be written as 
\begin{equation}
\begin{aligned} & \underset{\boldsymbol{x}}{\text{minimize}} &  & \frac{\boldsymbol{x}^{H}\boldsymbol{P}\boldsymbol{x}}{\boldsymbol{x}^{H}\boldsymbol{Q}\boldsymbol{x}}\\
 & \text{subject to} &  & \left|x_{i}\right|=1,
\end{aligned}
\label{eq:prob_x}
\end{equation}
where $\boldsymbol{P}=\sum_{k=1}^{K}\zeta_k \boldsymbol{J}_{k}^{H}\boldsymbol{w}\boldsymbol{w}^{H}\boldsymbol{J}_{k}+\frac{\sigma^{2}}{N}\Vert\boldsymbol{w}\Vert^{2}$
and $\boldsymbol{Q}=\boldsymbol{w}\boldsymbol{w}^{H}$ with $\boldsymbol{w}=[\widetilde{\boldsymbol{w}}]_{M+1:M+N}$.

Through slack variables, problem (\ref{eq:prob_x}) is equivalent to 
\begin{equation}
\begin{aligned} & \underset{\boldsymbol{x},\boldsymbol{y},\boldsymbol{z}}{\text{minimize}} &  & \frac{\boldsymbol{y}^{H}\boldsymbol{P}\boldsymbol{y}}{\boldsymbol{z}^{H}\boldsymbol{Q}\boldsymbol{z}}\\
 & \text{subject to} &  & \left|x_{i}\right|=1,\boldsymbol{x}=\boldsymbol{y}=\boldsymbol{z}.
\end{aligned}
\end{equation}
Its Lagrangian function is 
\begin{equation}\label{eq:Lagrangian}
\footnotesize{\mathcal{L}\left(\boldsymbol{x},\boldsymbol{y},\boldsymbol{z},\boldsymbol{u},\boldsymbol{v}\right)=\frac{\boldsymbol{y}^{H}\boldsymbol{P}\boldsymbol{y}}{\boldsymbol{z}^{H}\boldsymbol{Q}\boldsymbol{z}}+\frac{\rho_{1}}{2}\Vert\boldsymbol{y}-\boldsymbol{x}+\boldsymbol{u}\Vert^{2}+\frac{\rho_{2}}{2}\Vert\boldsymbol{z}-\boldsymbol{x}+\boldsymbol{v}\Vert^{2}.}
\end{equation}
At the $\ell$-th iteration of ADMM, the update rules are
\begin{equation}
\begin{cases}
\boldsymbol{x}_{\ell+1}=\arg\underset{\left|x_{i}\right|=1}{\min}\mathcal{L}\left(\boldsymbol{x},\boldsymbol{y}_{\ell},\boldsymbol{z}_{\ell},\boldsymbol{u}_{\ell},\boldsymbol{v}_{\ell}\right) & (a)\\
\boldsymbol{y}_{\ell+1}=\arg\underset{\boldsymbol{y}}{\min}~\mathcal{L}\left(\boldsymbol{x}_{\ell+1},\boldsymbol{y},\boldsymbol{z}_{\ell},\boldsymbol{u}_{\ell},\boldsymbol{v}_{\ell}\right) & (b)\\
\boldsymbol{z}_{\ell+1}=\arg\underset{\boldsymbol{z}}{\min}~\mathcal{L}\left(\boldsymbol{x}_{\ell+1},\boldsymbol{y}_{\ell+1},\boldsymbol{z},\boldsymbol{u}_{\ell},\boldsymbol{v}_{\ell}\right) & (c)\\
\boldsymbol{u}_{\ell+1}=\boldsymbol{u}_{\ell}+\boldsymbol{y}_{\ell+1}-\boldsymbol{x}_{\ell+1} & (d)\\
\boldsymbol{v}_{\ell+1}=\boldsymbol{v}_{\ell}+\boldsymbol{z}_{\ell+1}-\boldsymbol{x}_{\ell+1}. & (e)
\end{cases}\label{eq:ADMM_rules}
\end{equation}
In the following, we focus on solving (\ref{eq:ADMM_rules})-(a)(b)(c).
For notation simplicity, we ignore the subscript $\ell$ and $\ell+1$. 

For problem (\ref{eq:ADMM_rules})(a), it becomes 
\begin{equation}
\begin{aligned} & \underset{\boldsymbol{x}}{\text{minimize}} &  & \boldsymbol{x}^{H}\boldsymbol{x}-\text{Re}\left(\boldsymbol{a}^{H}\boldsymbol{x}\right)\\
 & \text{subject to} &  & \left|x_{i}\right|=1,
\end{aligned}
\label{eq:subprob_x}
\end{equation}
where $\boldsymbol{a}=\frac{2}{\rho_{1}+\rho_{2}}\left[\rho_{1}\left(\boldsymbol{y}+\boldsymbol{u}\right)+\rho_{2}\left(\boldsymbol{z}+\boldsymbol{v}\right)\right]$.
Since $\boldsymbol{x}^{H}\boldsymbol{x}=N$, the optimal solution
to problem (\ref{eq:subprob_x}) is $\boldsymbol{x}=e^{j\arg\left(\boldsymbol{a}\right)}$.

For problem (\ref{eq:ADMM_rules})(b), it can be further written as
\begin{equation}
    \begin{aligned} & \underset{\boldsymbol{y}}{\text{minimize}} &  & \boldsymbol{y}^{H}\tilde{\boldsymbol{P}}\boldsymbol{y}+\text{Re}\left(\boldsymbol{b}^{H}\boldsymbol{y}\right),
    \end{aligned}
\end{equation}
where $\tilde{\boldsymbol{P}}=\frac{\boldsymbol{P}}{\boldsymbol{z}^{H}\boldsymbol{Q}\boldsymbol{z}}+\frac{\rho_{1}}{2}\boldsymbol{I}\succ\boldsymbol{0}$
and $\boldsymbol{b}=\rho_{1}\left(\boldsymbol{u}-\boldsymbol{x}\right)$. The closed form solution is $\boldsymbol{y}=-\frac{1}{2}\tilde{\boldsymbol{P}}^{-1}\boldsymbol{b}$.

For problem (\ref{eq:ADMM_rules})(c), it becomes 
\begin{equation}
\underset{\boldsymbol{z}}{\text{minimize}}\quad\left(\boldsymbol{z}^{H}\tilde{\boldsymbol{Q}}\boldsymbol{z}\right)^{-1}+\frac{\rho_{2}}{2}\boldsymbol{z}^{H}\boldsymbol{z}+\text{Re}\left(\boldsymbol{c}^{H}\boldsymbol{z}\right)\label{eq:subprob_z}
\end{equation}
with $\tilde{\boldsymbol{Q}}=\frac{\boldsymbol{Q}}{\boldsymbol{y}^{H}\boldsymbol{P}\boldsymbol{y}}\succeq\boldsymbol{0}$
with rank 1 and $\boldsymbol{c}=\rho_{2}\left(\boldsymbol{v}-\boldsymbol{x}\right)$.
By letting $\tilde{\boldsymbol{Q}}=\boldsymbol{U}^{H}\boldsymbol{\Lambda}\boldsymbol{U}$
with $\boldsymbol{U}\boldsymbol{U}^{H}=\boldsymbol{U}^{H}\boldsymbol{U}=\boldsymbol{I}$, $\tilde{\boldsymbol{c}}=\boldsymbol{U}\boldsymbol{c}$,
and $\tilde{\boldsymbol{z}}=\boldsymbol{U}\boldsymbol{z}$, problem
(\ref{eq:subprob_z}) can be equivalently expressed as
\begin{equation}
\underset{\tilde{\boldsymbol{z}}}{\text{minimize}}\quad\left(\tilde{\boldsymbol{z}}^{H}\boldsymbol{\Lambda}\tilde{\boldsymbol{z}}\right)^{-1}+\frac{\rho_{2}}{2}\tilde{\boldsymbol{z}}^{H}\tilde{\boldsymbol{z}}+\text{Re}\left(\tilde{\boldsymbol{c}}^{H}\tilde{\boldsymbol{z}}\right).\label{eq:subprob_new_z}
\end{equation}
Since $\text{rank}\left(\tilde{\boldsymbol{Q}}\right)=1$, $\boldsymbol{\Lambda}=\text{diag}\left(\left\{ \lambda_{i}\right\} _{i=1}^{N}\right)$
with $\lambda_{1}=\frac{\Vert\boldsymbol{w}\Vert^{2}}{\boldsymbol{y}^{H}\boldsymbol{P}\boldsymbol{y}}$
and $\lambda_{2}=\cdots=\lambda_{N}=0$. Thus, the objective function of problem (\ref{eq:subprob_z}) can be expressed as 
\begin{equation}
\begin{aligned}f\left(\tilde{\boldsymbol{z}}\right)= &  \frac{1}{\lambda_1}\left|\tilde{z}_{1}\right|^{-2}+\frac{\rho_{2}}{2}\left|\tilde{z}_{1}\right|^{2}+\text{Re}\left(\tilde{c}_{1}^{*}\tilde{z}_{1}\right) \\
 & + \frac{\rho_{2}}{2}\hat{\boldsymbol{z}}^{H}\hat{\boldsymbol{z}}+\text{Re}\left(\hat{\boldsymbol{c}}^{H}\hat{\boldsymbol{z}}\right) ,
\end{aligned}
\end{equation}
where $\tilde{z}_{i}$ is the $i$-th element of $\tilde{\boldsymbol{z}}$,
$\hat{\boldsymbol{z}}=\left[\tilde{z}_{2},\ldots,\tilde{z}_{N}\right]^{T}$
and $\hat{\boldsymbol{c}}=\left[\tilde{c}_{2},\ldots,\tilde{c}_{N}\right]^{T}$. 

Therefore, problem (\ref{eq:subprob_new_z}) can be decomposed into
two independent problems. The first problem is as follows:
\begin{equation}
\underset{\tilde{z}_{1}}{\text{minimize}}\quad\left(\lambda_{1}\left|\tilde{z}_{1}\right|^{2}\right)^{-1}+\frac{\rho_{2}}{2}\left|\tilde{z}_{1}\right|^{2}+\text{Re}\left(\tilde{c}_{1}^{*}\tilde{z}_{1}\right).\label{eq:subprob_z_1}
\end{equation}
Since $\text{Re}\left(\tilde{c}_{1}^{*}\tilde{z}_{1}\right)\ge-\left|\tilde{c}_{1}\right|\left|\tilde{z}_{1}\right|$
with the equality achieved when $\arg\left(\tilde{z}_{1}\right)=\pi+\arg\left(\tilde{c}_{1}\right)$.
Then, solving problem (\ref{eq:subprob_z_1}) is essentially solving
\begin{equation}
\underset{\tilde{z}_{1}}{\text{minimize}}\quad\left(\lambda_{1}\left|\tilde{z}_{1}\right|^{2}\right)^{-1}+\frac{\rho_{2}}{2}\left|\tilde{z}_{1}\right|^{2}-\left|\tilde{c}_{1}\right|\left|\tilde{z}_{1}\right|.
\end{equation}
Let $t=\left|\tilde{z}_{1}\right|\ge0$ and $g\left(t\right)=\frac{1}{\lambda_{1}t^{2}}+\frac{\rho_{2}}{2}t^{2}-\left|\tilde{c}_{1}\right|t$,
then the first-order optimality condition is 
\begin{equation}
\rho_{2}t^{4}-\left|\tilde{c}_{1}\right|t^{3}-\frac{2}{\lambda_{1}}=0.
\end{equation}
After obtaining the positive root of the quartic polynomial, we can
select the root with the largest value of $g\left(t\right)$. Hence,
the optimal solution is $\tilde{z}_{1}=-te^{j\arg\left(\tilde{c}_{1}\right)}$.

The second problem is as follows:
\begin{equation}
\underset{\hat{\boldsymbol{z}}}{\text{minimize}}\quad\frac{\rho_{2}}{2}\hat{\boldsymbol{z}}^{H}\hat{\boldsymbol{z}}+\text{Re}\left(\hat{\boldsymbol{c}}^{H}\hat{\boldsymbol{z}}\right),
\end{equation}
which has a closed form solution $\hat{\boldsymbol{z}}=-\frac{1}{\rho_{2}}\hat{\boldsymbol{c}}$.
Finally, the optimal solution of problem (\ref{eq:subprob_new_z})
is 
\begin{equation}
\tilde{\boldsymbol{z}}=\left[-te^{j\arg\left(\tilde{c}_{1}\right)},-\frac{\tilde{c}_{2}}{\rho_{2}},\ldots,-\frac{\tilde{c}_{N}}{\rho_{2}}\right]^{T}.
\end{equation}
The optimal solution to problem (\ref{eq:subprob_z}) is $\boldsymbol{z}=\boldsymbol{U}^{H}\tilde{\boldsymbol{z}}$.

In a summary, the derived algorithm will be double-loop, where the outer loop accounts for the AO between $\widetilde{\boldsymbol{w}}$ and $\widetilde{\boldsymbol{x}}$, and the inner loop accounts for the ADMM of $\widetilde{\boldsymbol{x}}$.

\begin{figure}[t]
    \captionsetup{aboveskip=0pt}
    \centering
    \includegraphics[width=.8\linewidth]{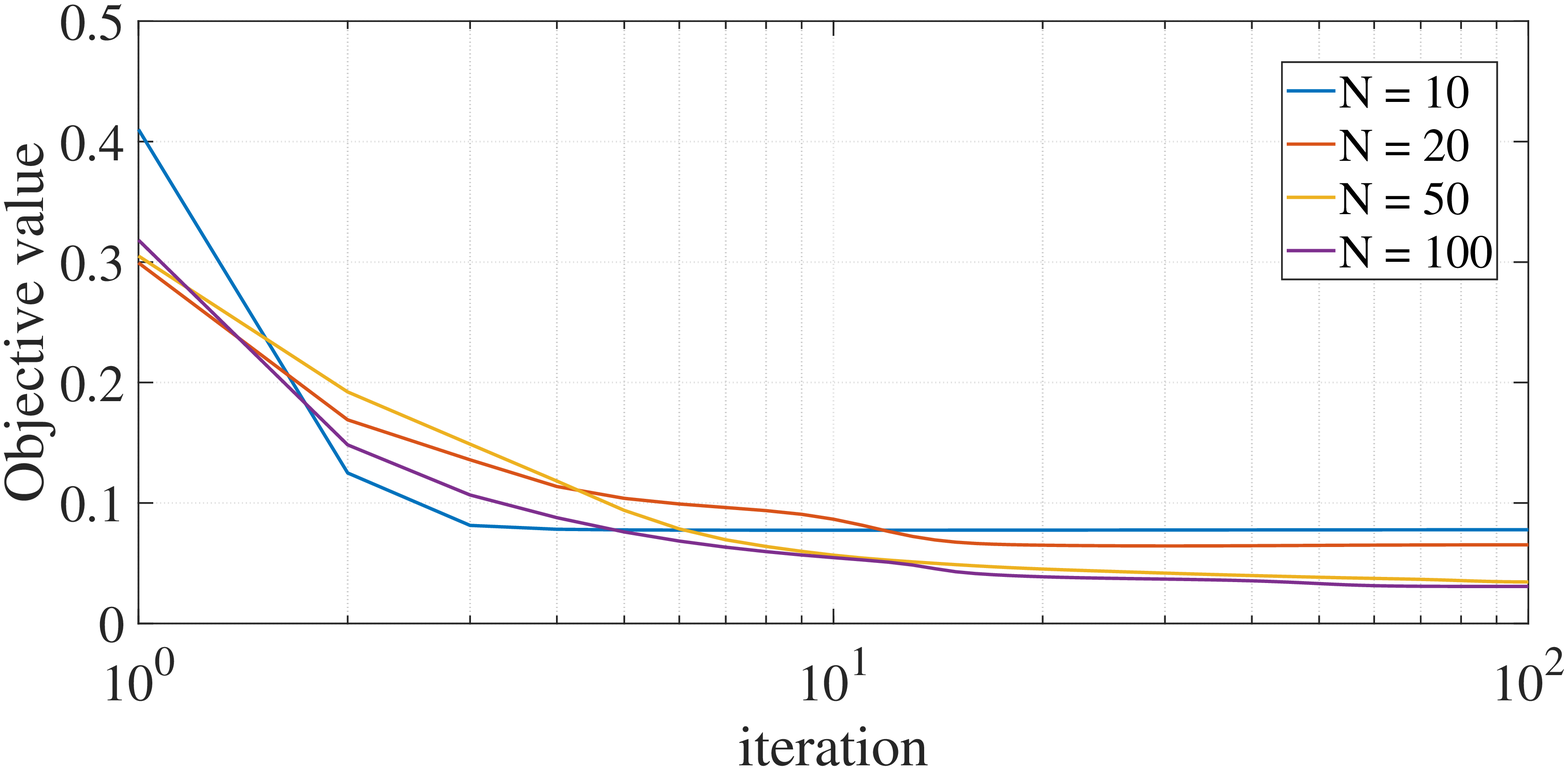}
    \caption{Convergence of the objective function, where $M = 0$.}
    \label{fig:Conv}
\end{figure}
\begin{figure}[t]
\captionsetup{aboveskip=0pt}
\captionsetup[subfigure]{aboveskip=-5pt}
    \centering
    \begin{subfigure}{.2\textwidth}
        \centering
        \includegraphics[width=\linewidth]{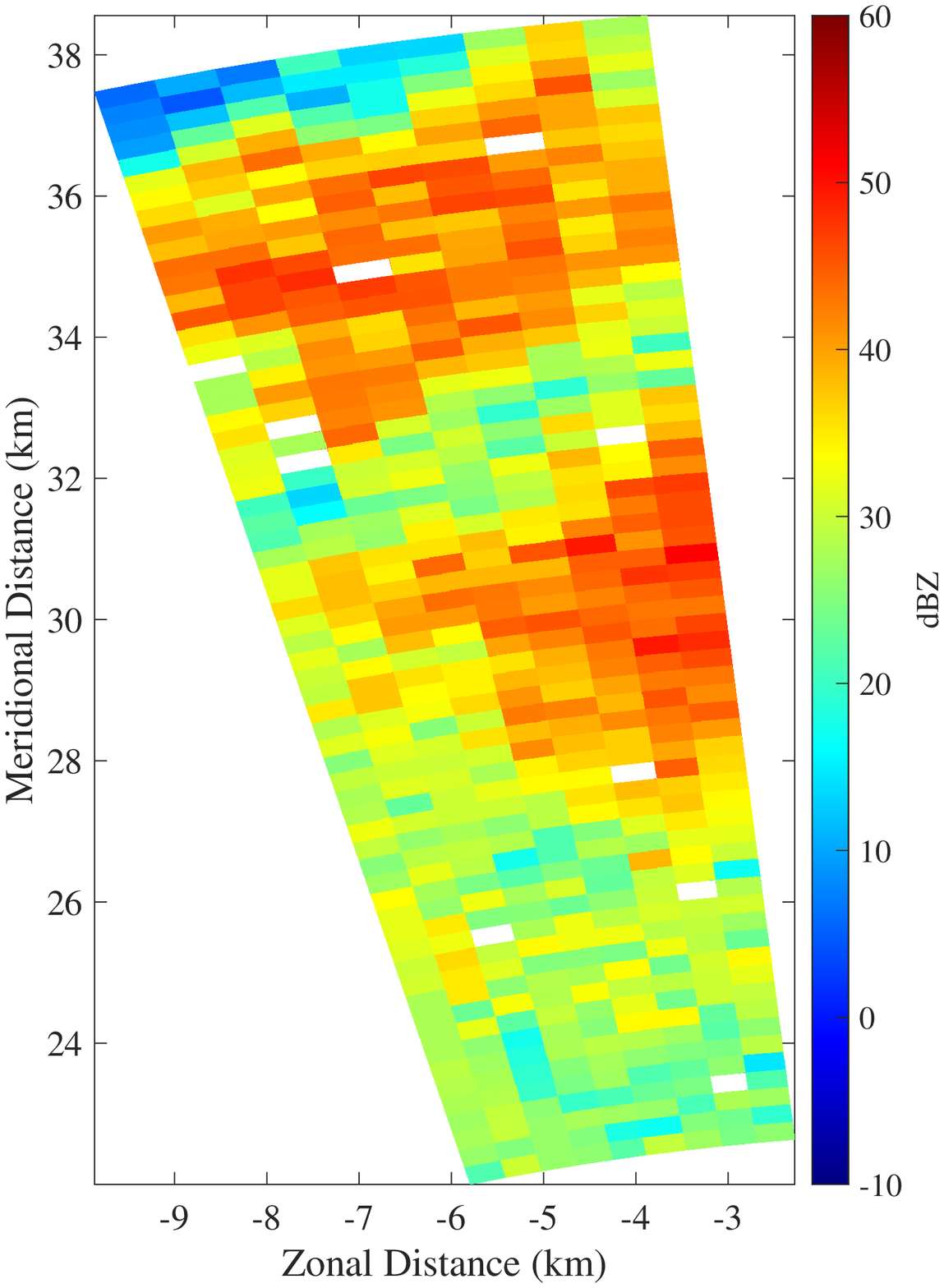}
        \caption{$Z$ (real data).}
    \end{subfigure}
    \begin{subfigure}{.2\textwidth}
        \centering
        \includegraphics[width=\linewidth]{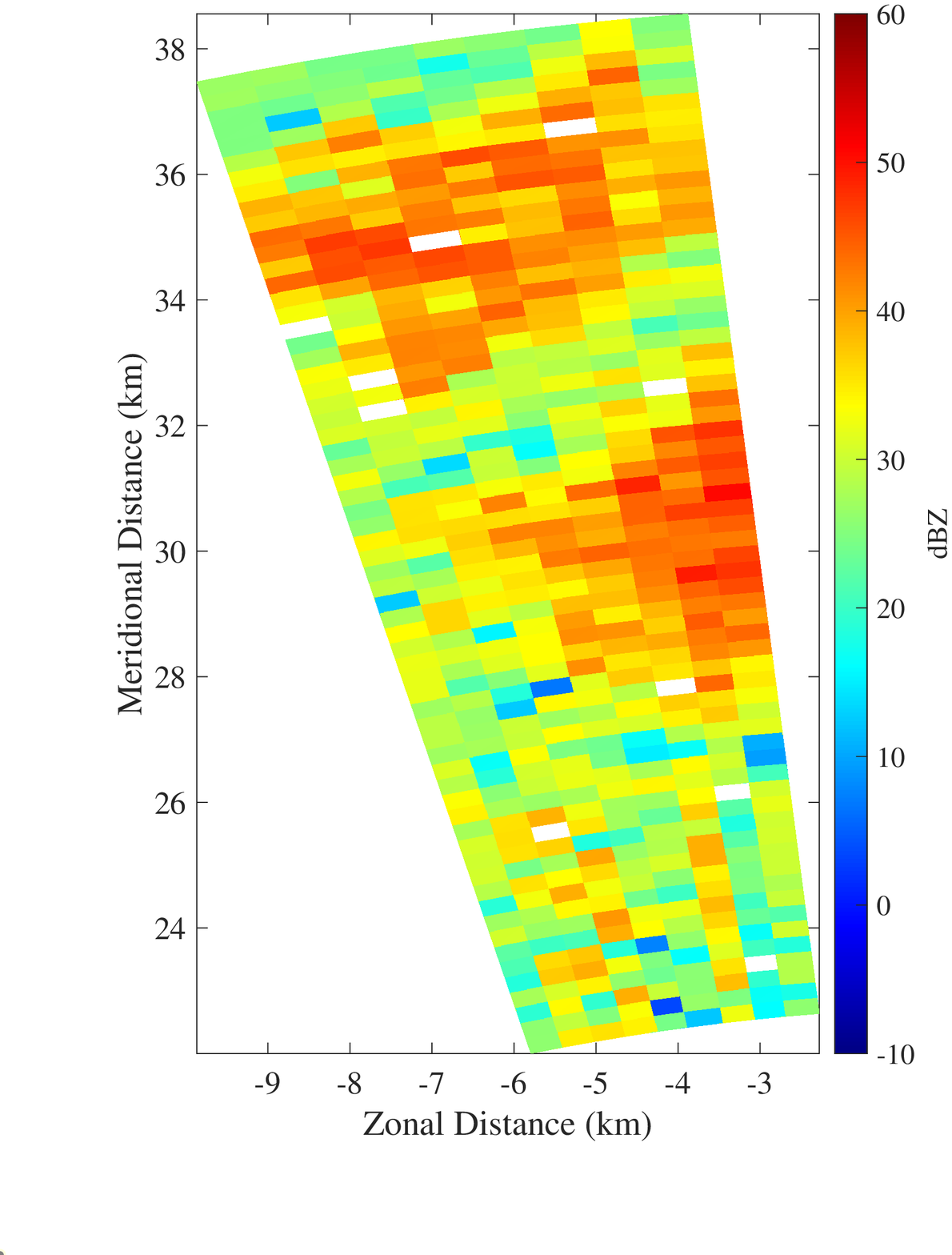}
        \caption{$Z$ (proposed method).}
    \end{subfigure}
    \begin{subfigure}{.2\textwidth}
        \centering
        \includegraphics[width=\linewidth]{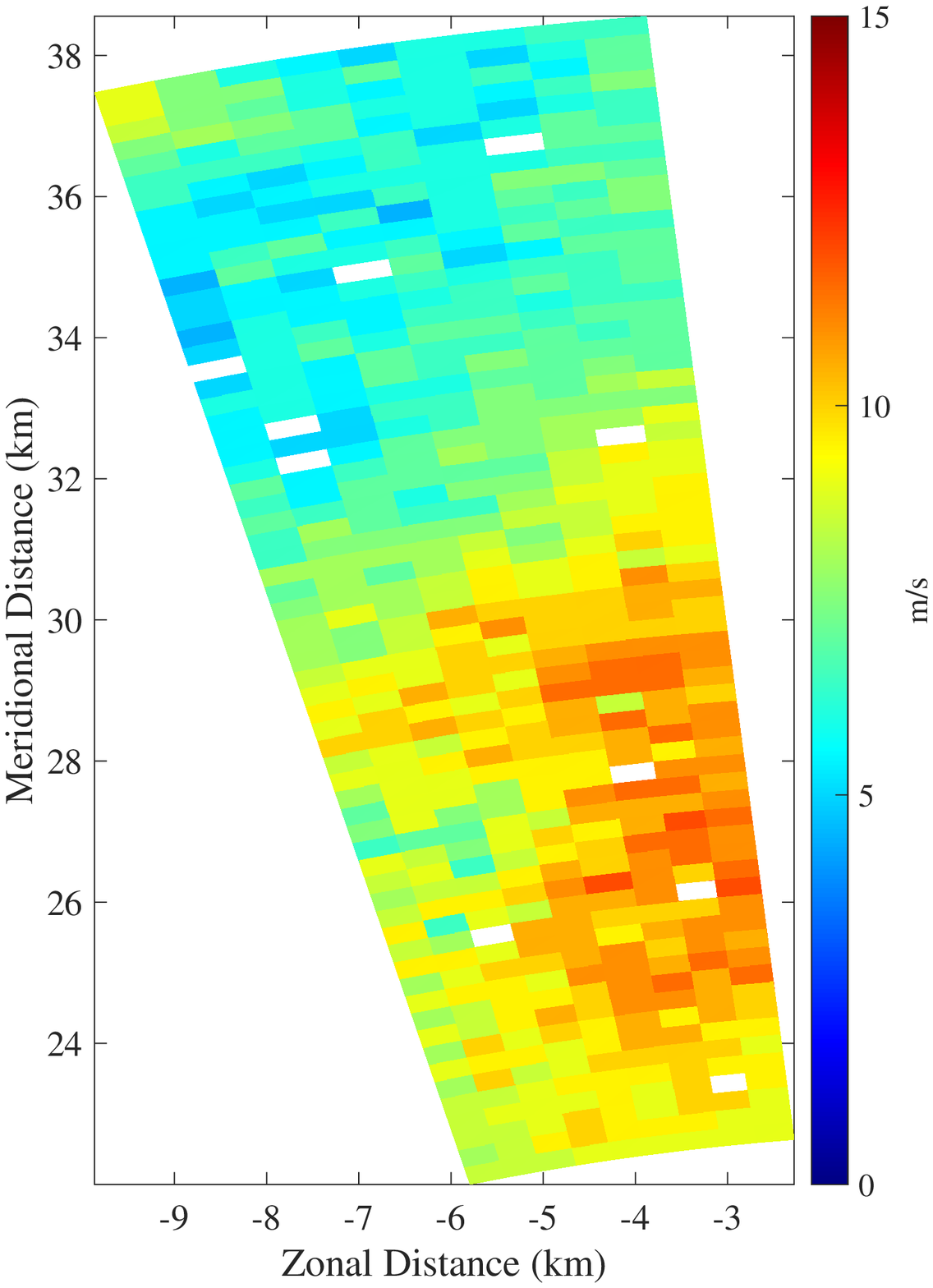}
        \caption{$V_{r}$ (real data).}
    \end{subfigure}
    \begin{subfigure}{.2\textwidth}
        \centering
        \includegraphics[width=\linewidth]{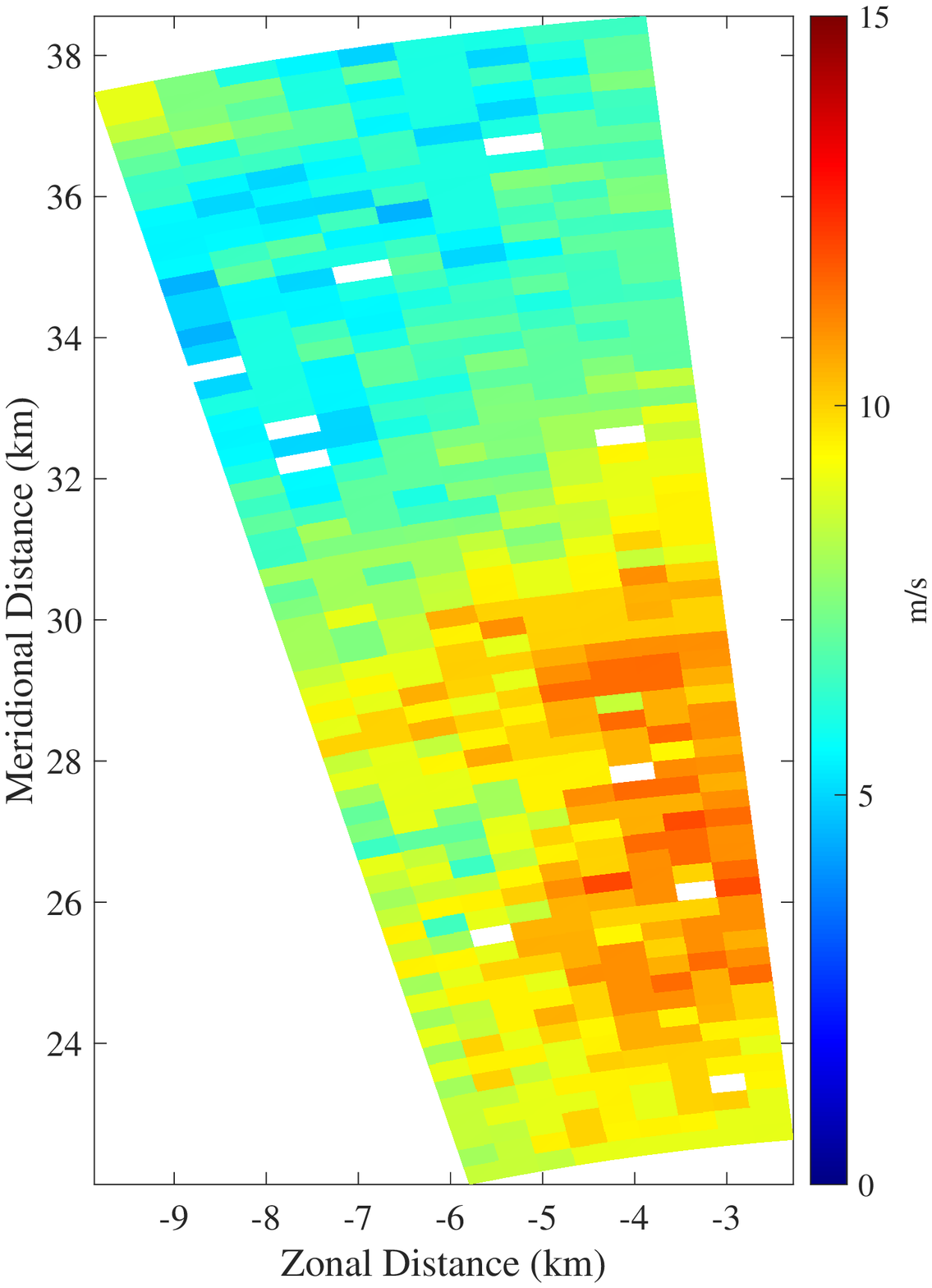}
        \caption{$V_{r}$ (proposed method).}
    \end{subfigure}   
    \caption{Consistency evaluation on reflectivity and radial velocity between real data and the designed  pulse compression.}
    \label{fig:RefZh}
\end{figure}

\section{Experiment Results}
\vspace{-0.1cm}
In this part, we study the performance of the optimized subpulse-filter pairs in weather observation. 
We conduct our simulation using a radar simulator \cite{MatlabWeatherToolbox}, which generates time series data for the NEXRAD  WSR88D radar specifications\footnote{The polarimetric weather radar NEXRAD WSR88D operates in $2.8$ GHz and has a maximum unambiguous range of $100$ km with $250$m resolution. 
} \cite{NexRADWSR88D}.
Following the same procedure as suggested by \cite{kumar2020intrapulse}, we convolve the simulated weather echoes with the transmit subpluses, which imposes the effect of pulse compression waveform. 
Next, we convolve the obtained uncompressed samples with the filter coefficients. Hence, we assess the utility of the optimized subpluse-filter pair based on Level-II information from NEXRAD  WSR88D.  

Fig. \ref{fig:Conv} shows the convergence of the proposed algorithm, from which we observe that the objective value decreases monotonically as the iteration goes.



In \figurename{~\ref{fig:RefZh}}, we set $N = \widetilde{N} = 100$ and evaluate the horizontal reflectivity\footnote{The reflectivity $Z$ (in units of $mm^6/m^3$) commonly span many orders of magnitude, and hence a logarithmic scale $\text{dB}Z=10\log_{10} Z$ is used.} and radial velocity\footnote{Radial velocity refers to the first moment of the power-normalized spectra, which reflects the air motion toward or away from the radar.} for real data and the simulated time series utilizing the optimized subpulse-filter pairs. It can be observed that the simulated result mimic the behaviour of the real data, which indicates the efficacy of utilizing the designed pulse compression.

\begin{figure}[t]
\captionsetup{aboveskip=-2pt}
\begin{centering}
\includegraphics[width=.2\textwidth]{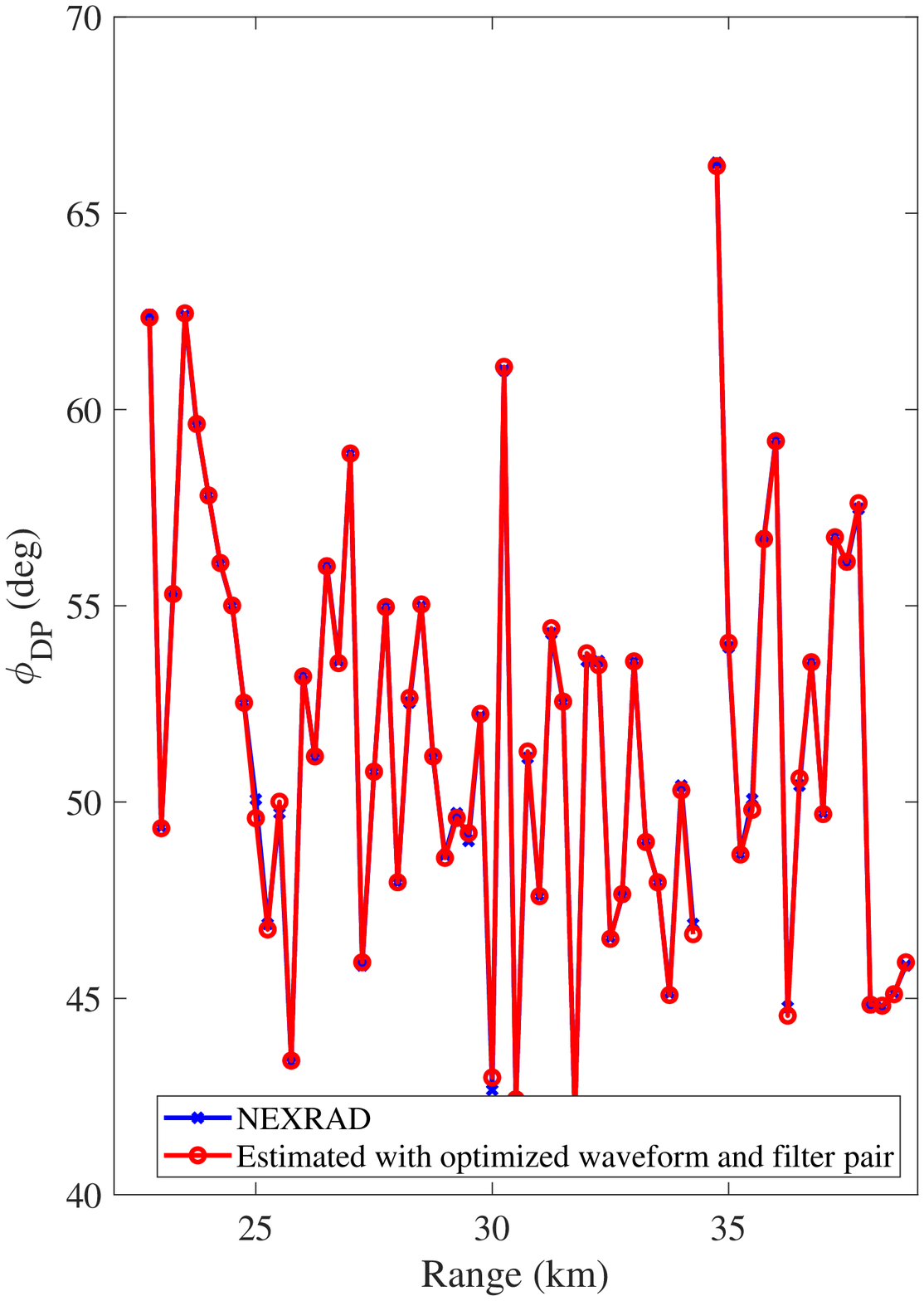}
\includegraphics[width=.2\textwidth]{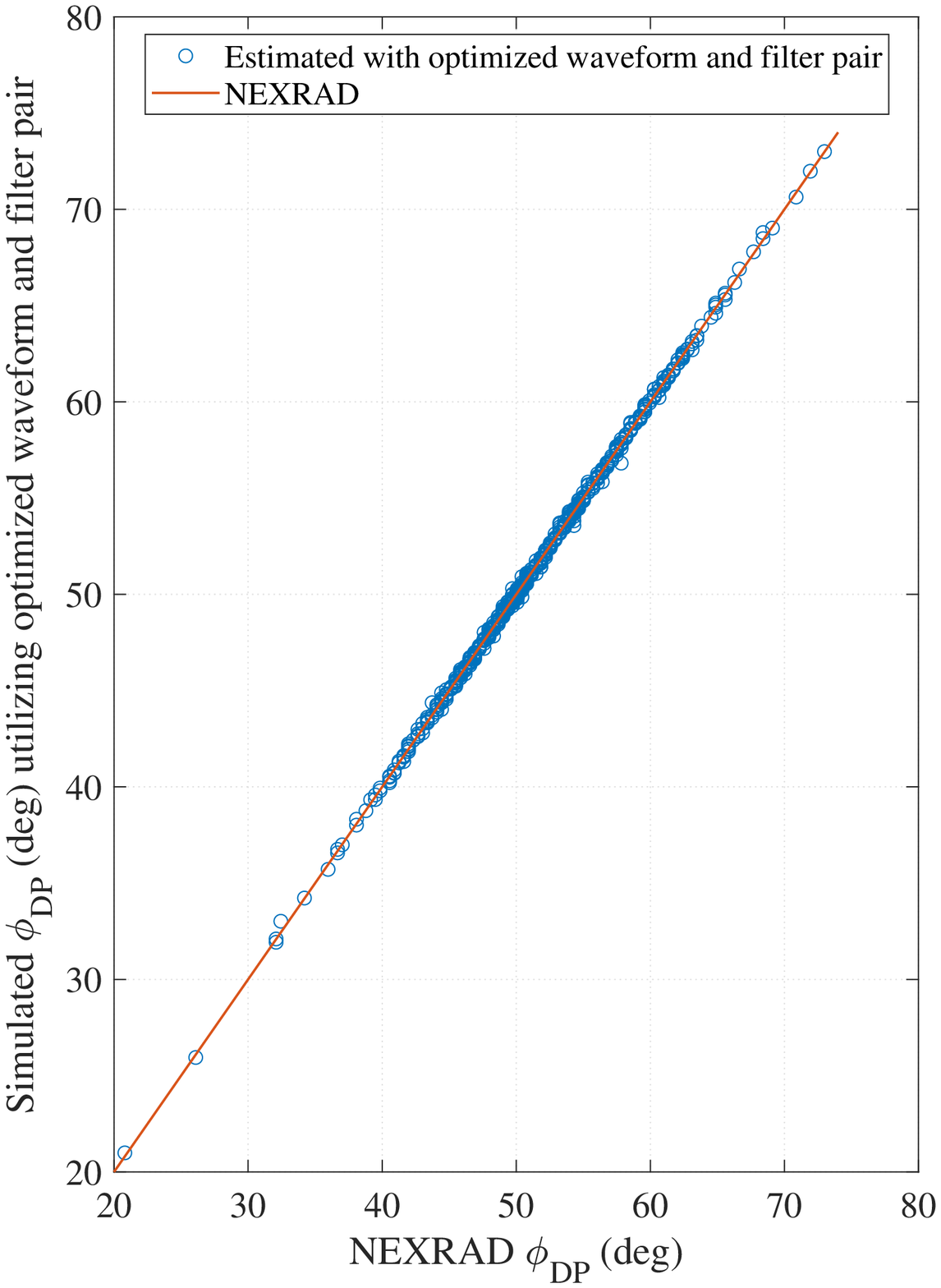}
\par\end{centering}
\caption{Differential phase of returned signal from the benchmark and proposed approach. Left: Range profile. Right: Consistency evaluation.}
\label{fig:PhiDR}
\end{figure}

\figurename{~\ref{fig:PhiDR}} indicates the difference in the phase delay ($\phi_{DP}$) of the returned signal
from the benchmark and simulated data by using the proposed method. The values of $\phi_{DP}$ provides information on the nature of the scatterers that are being sampled.
By these comparisons, it can be observed that the radar moment estimation is consistent to the NEXRAD specifications, which indicates high quality of the optimized subpulse-filter pairs in weather radar applications.

\section{Conclusion}
\label{sec:majhead}
We propose an optimization-based method for implementing a delicate pulse compression on weather radar. Through the proposed method, the transmitted subpulse codes and the extended filter are jointly designed, which guarantees that the pulse compression will possess a low sidelobe level and further a better estimation of the meteorological reflectivity. The experiments result demonstrates the efficacy of the proposed approach and the improvement brought by this delicately designed pulse compression technique.

\bibliographystyle{IEEEbib}
\bibliography{strings}

\begin{thebibliography}{10}

\bibitem{bc}
V.~N. Bringi and V~Chandrasekar,
\newblock {\em Polarimetric Doppler Weather Radar: Principles and
  Applications},
\newblock Cambridge University Press, 2001.

\bibitem{fetter1970radar}
R.~W. Fetter,
\newblock ``Radar weather performance enhanced by pulse compression,''
\newblock in {\em Proceedings of 14th AMS Conf. Radar Meteorol}, 1970, pp.
  413--418.

\bibitem{mudukutore1998pulse}
A.~S. Mudukutore, V~Chandrasekar, and R.~J. Keeler,
\newblock ``Pulse compression for weather radars,''
\newblock {\em IEEE Transactions on geoscience and remote sensing}, vol. 36,
  no. 1, pp. 125--142, 1998.

\bibitem{george2008considerations}
J.~George, N.~Bharadwaj, and V~Chandrasekar,
\newblock ``Considerations in pulse compression design for weather radars,''
\newblock in {\em IGARSS 2008-2008 IEEE International Geoscience and Remote
  Sensing Symposium}. IEEE, 2008, vol.~5, pp. V--109.

\bibitem{beauchamp2016pulse}
R.~M. Beauchamp, S.~Tanelli, E.~Peral, and V~Chandrasekar,
\newblock ``Pulse compression waveform and filter optimization for spaceborne
  cloud and precipitation radar,''
\newblock {\em IEEE Transactions on Geoscience and Remote Sensing}, vol. 55,
  no. 2, pp. 915--931, 2016.

\bibitem{kumar2020intrapulse}
M.~Kumar and V~Chandrasekar,
\newblock ``Intrapulse polyphase coding system for second trip suppression in a
  weather radar,''
\newblock {\em IEEE Transactions on Geoscience and Remote Sensing}, vol. 58,
  no. 6, pp. 3841--3853, 2020.

\bibitem{song2015optimization}
J.~Song, P.~Babu, and D.~P. Palomar,
\newblock ``Optimization methods for designing sequences with low
  autocorrelation sidelobes,''
\newblock {\em IEEE Transactions on Signal Processing}, vol. 63, no. 15, pp.
  3998--4009, 2015.

\bibitem{kerahroodi2017coordinate}
M.~A. Kerahroodi, A.~Aubry, A.~De~Maio, M.~M. Naghsh, and M.~Modarres-Hashemi,
\newblock ``A coordinate-descent framework to design low psl/isl sequences,''
\newblock {\em IEEE Transactions on Signal Processing}, vol. 65, no. 22, pp.
  5942--5956, 2017.

\bibitem{boyd2011distributed}
S.~Boyd, N.~Parikh, and E.~Chu,
\newblock {\em Distributed optimization and statistical learning via the
  alternating direction method of multipliers},
\newblock Now Publishers Inc, 2011.

\bibitem{4567600}
P.~Stoica, J.~Li, and M.~Xue,
\newblock ``On binary probing signals and instrumental variables receivers for
  radar,''
\newblock {\em IEEE Transactions on Information Theory}, vol. 54, no. 8, pp.
  3820--3825, 2008.

\bibitem{4787625}
U.~Niesen, D.~Shah, and G.~W. Wornell,
\newblock ``Adaptive alternating minimization algorithms,''
\newblock {\em IEEE Transactions on Information Theory}, vol. 55, no. 3, pp.
  1423--1429, 2009.

\bibitem{MatlabWeatherToolbox}
``Simulating a polarimetric radar return for weather observation,''
  \url{https://nl.mathworks.com/help/radar/ug/simulating-a-polarimetric-radar-return-for-weather-observation.html},
\newblock Accessed: 2021-12-16.

\bibitem{NexRADWSR88D}
``Nexrad technical information,''
  \url{https://www.roc.noaa.gov/WSR88D/Engineering/NEXRADTechInfo.aspx},
\newblock Accessed: 2021-12-16.

\end{thebibliography}

\end{document}